\documentclass[pra,aps,preprint,showpacs]{revtex4-2}
\usepackage{amssymb}
\usepackage{amsmath}
\usepackage{amsfonts}
\usepackage{graphicx}
\usepackage{color}
\usepackage{bm}

\usepackage[sort&compress]{natbib}
\begin{document}

\title{Phase delays in $\omega-2\omega$ above-threshold ionization}

\author{S. D. L\'{o}pez$^1$, S. Donsa$^2$, S. Nagele$^2$, D. G. Arb\'{o}$^{1,3}$ and J. Burgd\"{o}rfer$^2$}

\affiliation{$^1$ Institute for Astronomy and Space Physics - IAFE
(CONICET-UBA), CC 67, Suc. 28, C1428ZAA, Buenos Aires, Argentina}
\affiliation{$^2$Institute for Theoretical Physics, Vienna University of Technology, Wiedner Hauptstr. 8-10/E136, A-1040 Vienna, Austria, EU}
\affiliation{$^3$Facultad de Ciencias Exactas y Naturales y Ciclo B\'asico Com\'un, Universidad de Buenos Aires, Argentina}

\begin{abstract}

Relative phases of atomic above-threshold ionization
wavepackets have been investigated in a recent experiment [L. J. Zipp, A. Natan, and P. H. Bucksbaum, Optica \textbf{1}, 361-364 (2014)] exploiting interferences between different pathways in a weak probe field at half the frequency of the strong ionization pulse.
In this work we theoretically explore the extraction of phase delays and time delays of attosecond wavepackets formed in strong-field ionization.
We perform simulations solving the time-dependent Schr\"odinger equation and compare these results with the strong-field and Coulomb-Volkov approximations.
In order to disentangle short- from long- ranged effects of the atomic potential we also perform simulations for atomic model potentials featuring a Yukawa-type short-range potential.
We find significant deviations of the \textit{ab-initio} phase delays between different photoelectron pathways from the predictions by the strong-field approximation even at energies well above the ionization threshold. We identify similarities but also profound differences to the well-known interferometric extraction of phase- and time delays in one-photon ionization.

\end{abstract}

\pacs{32.80.Rm,32.80.Fb,03.65.Sq}

\date{\today}

\maketitle

\section{Introduction}

Measuring and analyzing ionization phases and timing information on electron wavepackets ionized by absorption of an XUV photon represents one of the major advances attosecond pulses and phase-controlled femtosecond laser pulses have enabled during the last decade \cite{Krausz09,PazourekRMP15}.
Such XUV pulses in combination with near-infrared or visible (NIR/V) laser light permit the control of electronic motion on the shortest accessible timescales \cite{Veniard95,Schins96,Glover96,Hummert20}.
Pump-probe techniques such as attosecond streaking \cite{ItataniPRL02, Goulielmakis04,Goulielmakis08} and reconstruction of attosecond harmonic beating by interference of two-photon transitions (RABBIT) \cite{VeniardPRA1996,Paulsci01} have been employed to measure attosecond time-resolved electron emission from noble gas atoms \cite{SchultzeSci2010, klunderPRL2011, Guenot2012, guenot2014, Dalhstrom13, Fuchs20}, molecules \cite{Huppert16,Beaulieu17}, and solids \cite{Cavalieri07,Lemell15,Haessler15}.
Whereas attosecond streaking of electrons ionized by an XUV pulse can be understood in terms of a classical time-resolved shift in momentum and energy by the probing IR field  \cite{ItataniPRL02,NagelePRA12,PazourekPRL12,DellaPicca20a,DellaPicca20b,Dahlstrom}, RABBIT employs two interfering quantum paths to the same final state in the continuum called sideband \cite{klunderPRL2011,Dahlstrom}.
This sideband energy can be reached through a two-photon process involving absorption of photons from one of two adjacent harmonic orders of a high-order harmonic generation (HHG) radiation followed by absorption or emission of an IR photon of the fundamental driving frequency $\omega$ \cite{KheifetsPRA13,feist14,Su13}.

Two-color ($\omega - 2\omega$) laser fields with well-controlled relative phases between both colors have been experimentally and theoretically studied since the last decade of the last century \cite{Schumacher1994,Arbo2015,Ehlotzky2001,Xie2012,Arbo2014}. 
Recently, they have also been employed as alternative tool to extract information on ionization phases and time delays \cite{You20,Fuchs20s,Donsa19,Laurent12}.
One key feature is that the broken inversion symmetry of the $\omega - 2\omega$ field allows for interference between odd and even partial waves of the outgoing photoelectron which leads to a $(\theta \leftrightarrow \pi-\theta)$ asymmetry of the emission signal.

\begin{figure}[tbp]
	\includegraphics[width=10cm]{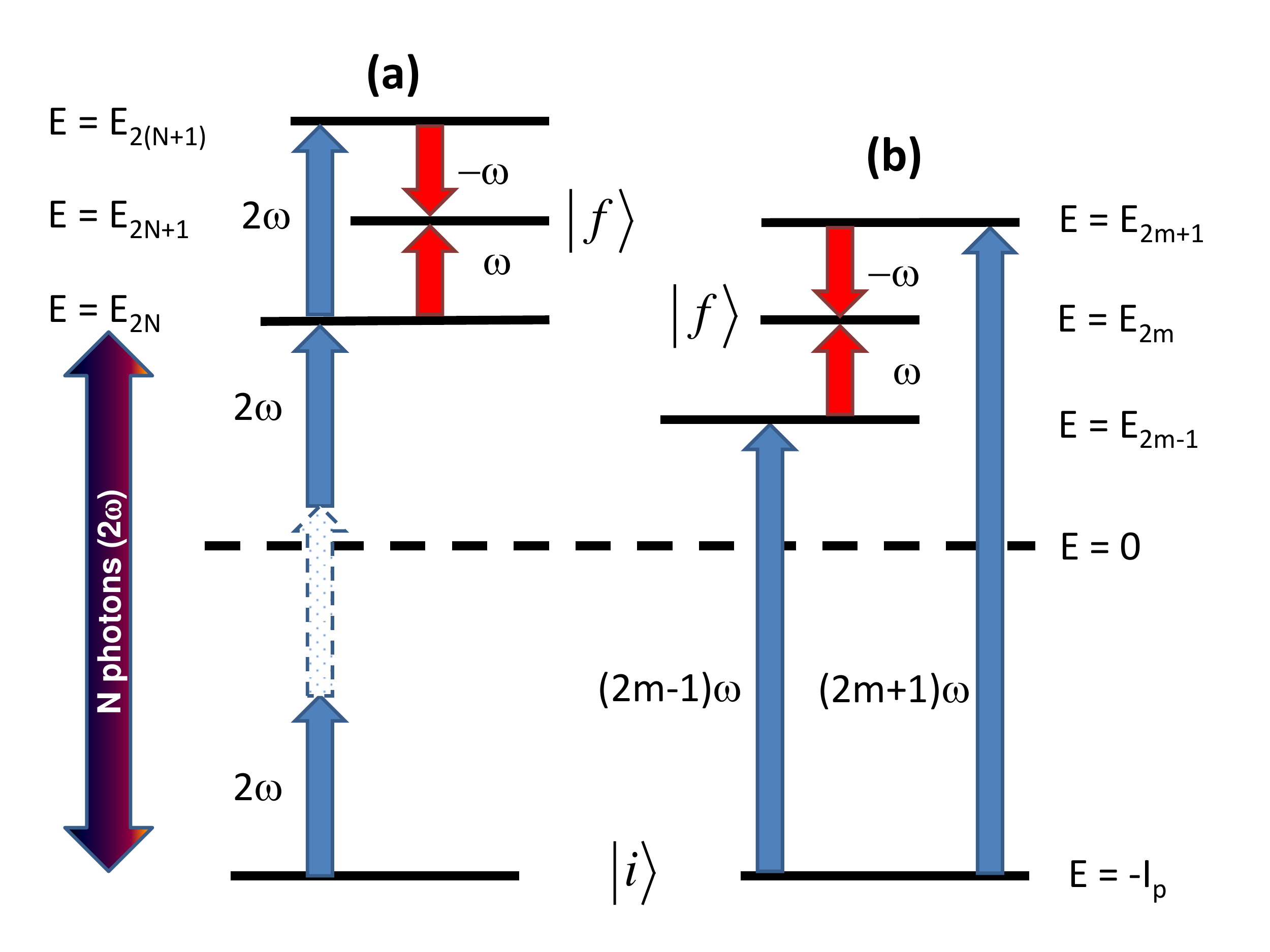}
	\caption{Comparison between (a) multi-photon strong-field interference (MPSFI) and the standard RABBIT protocol (b) for two interfering pathways from the initial bound state $ \left|i \right\rangle $  to final states $\left| f \right\rangle $ in the continuum, schematically. While RABBIT applies to two pathways involving ionization by one photon with energies $\left( 2m-1 \right) \omega$ and $\left( 2m+1 \right) \omega$ generated by HHG, MPSFI involves (at least) two ATI peaks generated by absorbing $N$ or $\left( N+1 \right)$ photons from the strong pump field with frequency $2\omega$. The final state $\left| f \right\rangle $ is reached in either case by the absorption ($\omega$) or emission ($-\omega$) of one photon of the weak probe field. Each arrow denotes a one-photon transition.}
\label{interference_sketch}
\end{figure}

Recently, Zipp \textit{et al.} \cite{Zipp} extended the measurement of ionization phases and attosecond time delays to the strong-field multiphoton regime, providing new perspectives on time-resolved strong-field ionization. In this novel $\omega - 2\omega$ interference protocol the role of electron wavepackets emitted by absorption of a single photon from either one or two subsequent harmonics in the RABBIT protocol is replaced by adjacent ATI peaks generated by a strong driving field of frequency $2\omega$.
The concomitant weaker $\omega$ field opens up interfering pathways to side bands in between neighboring ATI peaks by absorbing or emitting one $\omega$ photon. 
Measuring the photoelectron angular distribution as a function of the relative phase $\phi$ between the $\omega$ and the $2\omega$ fields provides information on the ATI amplitudes. This interferometric approach to multi-photon ionization (Fig.~\ref{interference_sketch}a) resembles the original RABBIT protocol for the extraction of the ionization phase in one-photon ionization (Fig.~\ref{interference_sketch}b). It promises new insights into relative phases and, possibly, attosecond-scale timing information of multi-photon strong-field processes.
Somewhat simplified, it can address the question what additional phase delays incur or how much longer it takes forming a wavepacket by absorbing $N+1$ rather than by $N$ photons. Some works based on the strong field approximation were recently reported in this direction \cite{Bertolino21,Feng19}.
Indeed, first simulations employing semiclassical trajectory methods \cite{Song18, Kheifets21,Feng19} highlighted the role of transient trapping of the wavepacket for the phase shift of the ATI peaks close to or even below the threshold.
A  detailed analysis of the information encoded in the ionization phases, their dependence on the intensities of the driving ($I_{2\omega}$) and probing ($I_{\omega}$) field, and on the properties of the atomic potential appears to be still missing. 

As will be shown in the following, multi-photon strong-field interference (MPSFI, Fig.~\ref{interference_sketch}a) substantially differs from the standard RABBIT protocol as a multitude of pathways with different number of photons and a broad range of partial waves of the emerging electronic wavepacket contribute. Phase delays can be extracted by this photoelectron interferometry not only at energies near the so-called sidebands but also near the ATI peaks. Moreover, in the strong-field setting  phase delays are, unlike in the RABBIT protocol, found to be remarkably sensitive to the probe field strengths, rendering the separation of the atomic field and laser field influence on the resulting phase and time delay more challenging.

In this work, we theoretically investigate the phase delays in the multi-photon regime accessible by such a $\omega - 2\omega$ interference protocol for two collinearly polarized laser fields. We find strong deviations of the time-dependent Schr\"odinger equation (TDSE) results from SFA predictions clearly indicating that the atomic potential has a crucial influence on the ionization phase of ATI peaks in this strong-field regime even at energies well above the ionization threshold. We also present a simplified analytical description of the MPSFI phase delays and discuss their potential to access timing information.

In Sec. \ref{methods} we briefly introduce the simulation methods employed. In Sec.~\ref{qpath} we present numerical results for quantum path interferences in multi-photon ionization. An approximate analytical approach to the extraction of the information on ionization phases, phase delays and time delays from such a $\omega - 2\omega$ protocol as well as numerical results for a model atom with a short-ranged Yukawa-type atomic binding potential are discussed in Sec.~\ref{analytical_model}. The comparison with experimental data for argon described by a suitable model potential \cite{Muller99}  in single active electron (SAE) approximation \cite{Arbo2015,Tong1997,Tong2000} is presented in Sec. \ref{results}. Concluding remarks are given in Sec. \ref{conclusions}. Atomic units are used unless stated otherwise.

\section{Methods}
\label{methods}

We consider a multi-femtosecond laser pulse with frequency $\omega$ and its second harmonic $2\omega$ with electric field amplitude
\begin{equation}
F(t)=f(t) \left[ F_{2\omega} \sin(2 \omega t+ \phi ) + F_{\omega} \sin(\omega t)\right] \hat{\bm{z}} ,
\label{field}
\end{equation}
where $f(t)$ is the overall pulse envelope and $\hat{\bm{z}}$ is the polarization direction of both fields. In the present $\omega - 2 \omega$ scenario $F_{2\omega}$ is the amplitude of the strong pump field giving rise to ATI peaks and $F_{\omega}$ is the amplitude of the weak probe field, i.e. $F_{\omega} \ll F_{2\omega}$. The relative phase $\phi$ between the $\omega$ and $2\omega$ fields is the experimentally accessible knob to control the interference between different multi-photon pathways. In the following, we will present results for the integral and angular differential photoelectron spectra as a function of $\phi$. For the envelope function we choose the form $f\left(t\right)=\sin^2\left(\frac{\pi t}{\tau}\right)$, where $\tau$ is the pulse duration covering 16 cycles in the strong pump field or eight cycles of the probe field, i.e., $\tau=16\pi/\omega$.

We solve the time-dependent Schr\"odinger equation (TDSE) in the single-active electron (SAE) approximation in the length gauge \cite{Arbo2015,Tong1997,Tong2000},
\begin{equation}
i \frac{\partial \psi (\vec{r},t)}{\partial t} = 
\left( \frac{p^{2}}{2} + V_a(r) + \vec{r} \cdot \vec{F}(t)  \right) \psi (\vec{r},t) .
\label{TDSE}
\end{equation}
In our simulation for argon to be compared with the experiment \cite{Zipp} we employ as atomic potential $V_a$ in Eq.~\eqref{TDSE} the Muller model potential \cite{Muller99}.
In order to delineate the role of short-ranged and long-ranged potentials we alternatively use a Yukawa-type atomic potential
\begin{equation}
V_a\left(r\right)=-\frac{b}{r}e^{-r/a} ,
\label{yukawa-potential}
\end{equation}
with charge parameter $b$ and the screening length $a$.

In addition to full solutions of the TDSE, we employ two popular versions of the distorted-wave Born approximation (DWBA) that allow to account for multi-photon and strong-field processes, namely the strong-field approximation (SFA) \cite{Keldysh64,Faisal73,Reiss80} and the Coulomb-Volkov approximation (CVA) \cite{Jain78}. Accordingly, the transition amplitude from an initial atomic state $\left|\phi _{i}(t)\right\rangle$ to a final state $\left|\varphi_{\vec{k}}\right\rangle$ with asymptotic momentum $\vec{k}$ in the continuum, i.e.,  $a_{\vec{k}}(\phi) = \lim_{t \rightarrow \infty} \left\langle \varphi_{\vec{k}} \right| \left. \psi (t) \right\rangle$ in the DWBA, is given by 
\begin{equation}
a(\vec{k},\varphi) = -i\int\limits_{-\infty }^{+\infty }dt\ \langle \chi
_{\vec{k}}^{\textsc{DW}}(t)|z\,F\,(t)\left\vert \phi _{i}(t)\right\rangle .  
\label{Tif}
\end{equation}
From Eq.~\eqref{Tif}, the SFA follows when the Volkov state is used as the distorted wave \cite{Keldysh64,Faisal73,Reiss80}
\begin{equation}
\chi _{\vec{k}}^{(\textsc{DW})-}(\vec{r},t)=%
\chi _{\vec{k}}^{(\textsc{V})-}(\vec{r},t)=\frac{\exp \mathbf{[}i(\vec{k}+\vec{A}%
)\cdot \vec{r}\mathbf{]}}{\left( 2\pi \right) ^{3/2}}\exp \left[
-i\int_{t}^{+\infty}dt^{\prime }\frac{(\vec{k}+\vec{A}(t^{\prime }))^{2}}{2}\right] \ .
\label{Volkov}
\end{equation}
The CVA results when approximating the distorted wave by a product of the Volkov solution and the Coulomb wave \cite{Jain78}
\begin{equation}
\chi _{\vec{k}}^{(\textsc{DW})-}(\vec{r},t)=%
\chi _{\vec{k}}^{(CV)-}(\vec{r},t) = \chi_{\vec{k}}^{(V)-}(\vec{r},t)\;\mathcal{D}_{C}(Z_{T},\vec{k},t),
\label{CV}
\end{equation}
where $\mathcal{D}_{C}(Z_{T},\vec{k},t)=N_{T}^{-}(k)\
_{1}F_{1}(-iZ_{T}/k,1,-ik\ r-i\vec{k}\cdot \vec{r})$ for a hydrogenic atom. The Coulomb normalization factor
$N_{T}^{-}(k)=\exp (\pi Z_{T}/2k)\Gamma (1+iZ_{T}/k)$
coincides with the amplitude of the Coulomb wave function at the origin, $%
_{1}F_{1}$ denotes the confluent hypergeometric function, and $Z_{T}$ is the
electric charge of the parent ion.
Eq. ~\eqref{Volkov} describes the final state of a free electron wave in the strong laser field while completely neglecting the atomic potential. The CVA in Eq.~\eqref{CV} includes also the Coulomb scattering of the free electron but neglects the effect of binding and of dynamical Stark shifts. These two DWBA approximations  provide points of reference for identifying dynamical multi-photon effects on ionization phases.

Because of the azimuthal symmetry, the electron probability distribution $P(\vec{k})=\left| a_{\vec{k}}\right|^2$ depends only on the electron momentum parallel ($k_{z}$) and transverse ($k_{\perp}$) to the field polarization direction or, alternatively, on the kinetic energy $E$ and the polar emission angle $\theta$, i.e., $P\left({k_{\perp},k_{z},\phi}\right)= (2 E)^{-1/2} P\left({E,\cos\theta,\phi}\right)$.
In the multiphoton regime, the photoelectron spectrum is composed of a series of peaks positioned at energies $E_n$
\begin{equation}
E_{n} = n \omega - (I_{p}+U_{p}),
\label{Econs}
\end{equation}
corresponding to absorption of a given number of $n_{\omega}$ photons of frequency $\omega$ and $N$ photons of frequency $2\omega$, such that $n \omega = n_{\omega} \omega + N  \left( 2 \omega \right)$. In Eq.~\eqref{Econs}, $I_{p}$ and $U_{p}$ denote the ionization potential and the ponderomotive energy, respectively.
As a given peak $E_n$ can be reached by different combinations of photon numbers $n_{\omega}$ and $N$, photo-electron interferometry in this strong-field setting is characterized by multi-path interferences of partial waves with opposite parity. Consequently, an important quantity for characterizing interferences between partial waves of opposite parity and, thus, to map out ionization phases in the $\omega - 2\omega$ protocol is the forward-backward ($\theta \leftrightarrow \pi - \theta$) asymmetry of the photoelectron emission probability
\begin{equation}
A(E,\phi)=\frac{S_{+}(E,\phi)-S_{-}(E,\phi)}{S_{+}(E,\phi)+S_{-}(E,\phi)} ,
\label{asymmetry}
\end{equation}
where the forward (backward) emission spectra $S_{+}$ ($S_{-}$) are obtained by integrating the momentum distribution over the +z (-z) hemisphere
\begin{equation}
S_{+(-)}(E,\phi)=\int_{0(-1)}^{1(0)} \mathrm{d}\cos\theta \: P(E,\cos\theta,\phi) .
\label{hemispheres}
\end{equation}

The calculated (or measured) signal function, generically denoted by $S(E,\phi)$, representing in the following either the photoemission probability into one hemisphere, $S_{+(-)}$ [Eq.~\eqref{hemispheres}] or the photoelectron asymmetry $A(E,\phi)$ [Eq.~\eqref{asymmetry}], can be written in terms of a Fourier series in the relative phase $\phi$. The emission signal takes the form \cite{Donsa20}
\begin{equation}
S(E,\phi)=c_{0}(E)+\sum_{i=1}^{\infty}c_{i}(E) \cos(i\phi-\delta_{i}(E)), 
\label{asym-cont}
\end{equation}
where the leading term ($i=1$) provides the information of the relative ionization phase $\delta_{1}(E) = \delta(E)$ in analogy to the RABBIT protocol \cite{Donsa20}. Higher-order Fourier components $c_{i}(E)$ for $\left\{ i=2,3...\right\}$ should provide an error estimate of the fit. Extending the analogy to RABBIT, Zipp \textit{et al.} \cite{Zipp} introduced an Eisenbud-Wigner-Smith (EWS) -type time delay \cite{Wigner55,Smith60} by mapping the phase delay $\delta (E)$ onto a time delay as
\begin{equation}
\tau(E)=\frac{\delta(E)}{2\omega}.
\label{delay-cont}
\end{equation}
Eq.~\eqref{delay-cont} can be viewed as finite-difference approximation to the spectral derivative $d\delta(E)/dE$ of the phase shift $\delta(E)$. We explore the physical significance of $\delta(E)$ and $\tau(E)$ in more detail below.

\section{Energy dependence of phase delay in multi-photon ionization}
\label{qpath}

\begin{figure}[tbp]
\includegraphics[width=10cm]{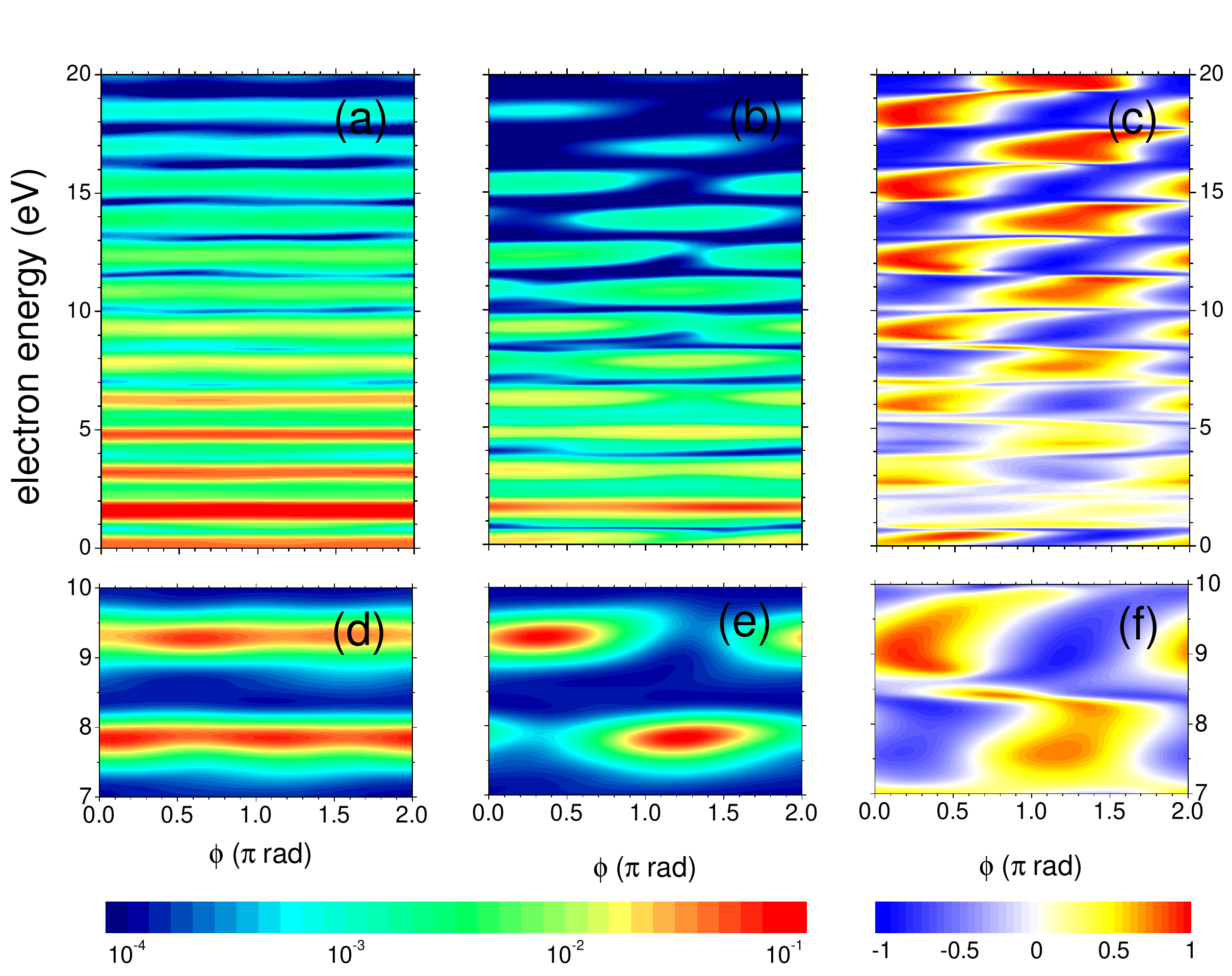}
\caption{(a) Total photoelectron spectrum (logarithmic scale), (b) forward emission spectrum integrated in the +z hemisphere (logarithmic scale) and (c) asymmetry parameter $S$ (linear scale) for argon as a function of the relative phase $\phi$ and the electron energy (in eV) calculated within the TDSE. The laser intensities are $I_{2\omega}=8\times10^{13}$ W/cm$^{2}$ and $I_{\omega}=4\times10^{11}$ W/cm$^{2}$ for the respective frequencies $2\omega$ and $\omega=0.057$ a.u. with pulse duration $\tau=881.85$ a.u., corresponding to eight full cycles of the latter. (d-f) Zoom in linear scale corresponding to (a-c), respectively.}
\label{ar-spectrum}
\end{figure}

As a representative example of $\omega - 2\omega$ atomic ionization, we choose the probe field with the fundamental frequency of a Ti:Sapphire laser of 800 nm wavelength in the near-infrared (NIR) region of the spectrum, and the pump frequency as its second harmonic with a 400 nm wavelength in the visible (V) region. In line with the experiment of Zipp \textit{et al.} \cite{Zipp} 
we study atomic ionization of argon by the two-color laser field in Eq. ~\eqref{field} with intensities $I_{2\omega}=c/ (8\pi) F_{2\omega}^{2}= 8\times10^{13}$ W/cm$^{2}$ and $I_{\omega}= c/ (8\pi) F_{\omega}^{2}=4\times10^{11}$ W/cm$^{2}$. In Fig. \ref{ar-spectrum} we exhibit the results of our TDSE calculations in the SAE approximation \cite{Tong1997,Tong2000}. In Fig. \ref{ar-spectrum}a we show the variation of the total multiphoton spectrum (integrated over all emission angles $\theta$) as a function of the relative two-color phase $\phi$. We observe the typical multiphoton peak structure with peak positions at energies predicted by Eq.~\eqref{Econs}, with the ionization potential for argon of $I_{p}=15.78$~eV and ponderomotive energy $U_{p}= 1.19$~eV. ATI peaks at even multiples of $\omega$ result predominantly from absorption of $N$ photons of frequency $2\omega$, while peaks at odd multiples of $\omega$ result from absorption or emission of at least one additional $\omega$ (probe) photon. Following the convention of RABBIT \cite{klunderPRL2011,Dahlstrom} we refer to the latter group of peaks with energies near odd multiples of the NIR frequency $\omega$
as ``sidebands'' (SB). Unlike the total electron emission integrated over all angles $\theta$ (Fig. \ref{ar-spectrum}a) whose $\phi$ dependence displays a $\pi$ periodicity (emission near $\phi$ and $\phi+\pi$ are identical), the emission into the forward hemisphere $S_+(E,\phi)$ given by Eq.~\eqref{hemispheres} ($0\leq\theta\leq\pi/2$) (Fig. \ref{ar-spectrum}b) and the asymmetry parameter $A(E,\phi)$ (Fig. \ref{ar-spectrum}c) display a $2\pi$ periodicity indicative of the parity-breaking contributions due to $\omega - 2\omega$  interferences. These structures are magnified in the close-up figures \ref{ar-spectrum}d, \ref{ar-spectrum}e, and \ref{ar-spectrum}f.  

\begin{figure}[tbp]
\includegraphics[width=10cm]{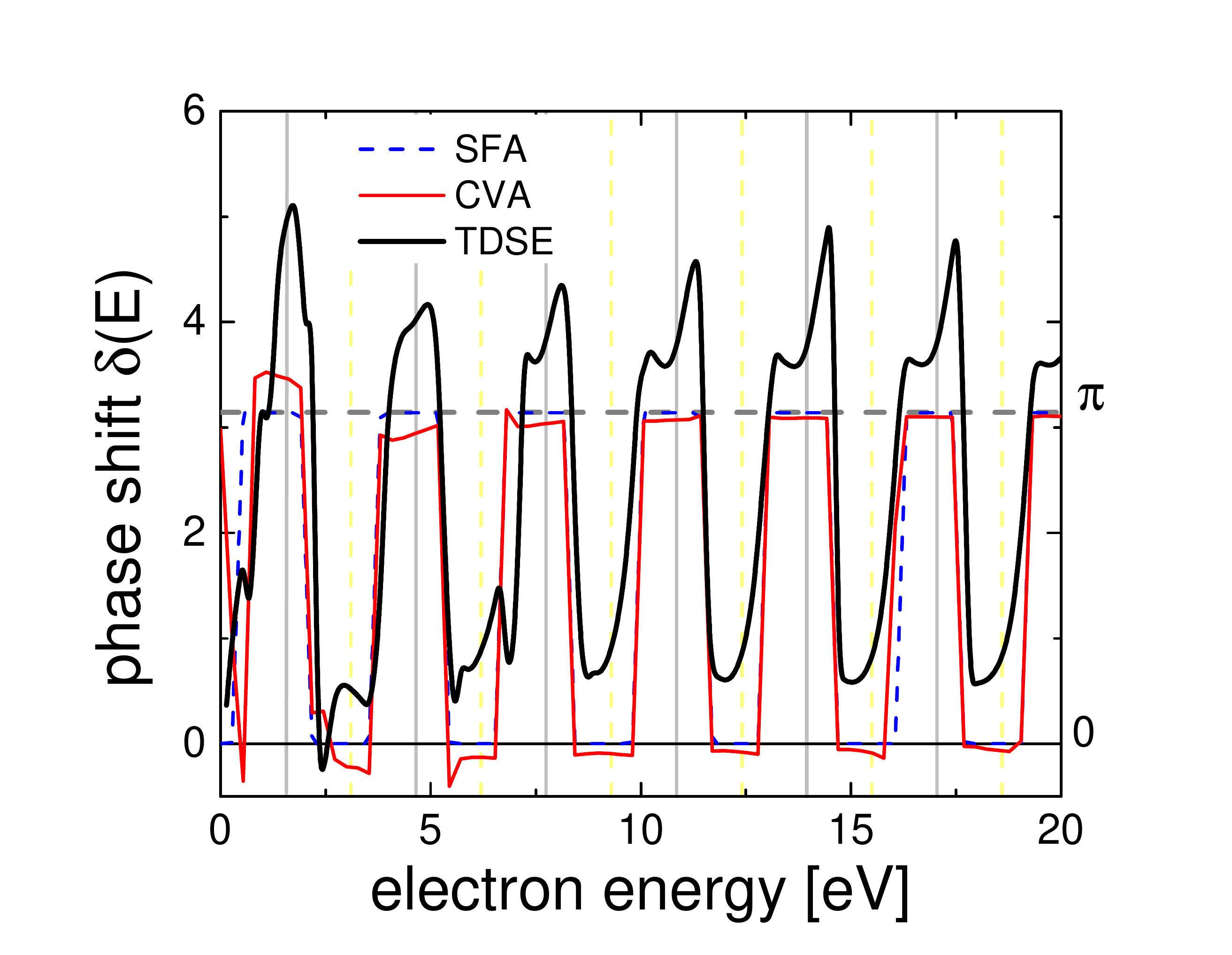}
\caption{Continuum phase shifts $\delta(E)$ extracted from the asymmetry $A(E,\phi)$ as a function of the emission energy from the TDSE (thick black solid line), SFA (dashed blue line), and CVA (thin red solid line) results. Thick solid vertical gray lines denote ATI peak energies and dashed vertical lines sideband energies according to Eq. (\ref{Econs}). The horizontal dashed line corresponds to the strong-field limit for ATI phase shifts [$\delta(E)=\pi$].}
\label{models-delays-cont}
\end{figure}

From the fit of the variation of the numerical data for $S_+(E,\phi)$ and $A(E,\phi)$ to the Fourier expansion [Eq. ~\eqref{asym-cont}] at fixed $E$, the relative ionization phase $\delta(E)$ can be extracted as the phase shift of the $\cos \phi$ oscillation. Because of the broad Fourier width of the ultrashort pulse [Eq.~\eqref{field}] the multiphoton electron spectrum (Fig.~\ref{ar-spectrum}a) is a continuous function of $E$. Accordingly, also the phase shift $\delta(E)$ can be viewed as a continuous function of $E$.
Results for the energy dependence of $\delta(E)$ predicted by the TDSE, the SFA, and the CVA calculations of $S(E,\phi)$ are shown in Fig. \ref{models-delays-cont}. Most strikingly, the SFA jumps almost discontinuously and periodically between $\pi$ near the ATI energies [even $n$ in Eq. (\ref{Econs})] and $0$ in the vicinity of side bands [odd $n$ in Eq. (\ref{Econs})]. The CVA introduces modest variations to this SFA behavior which are a signature of Coulomb scattering of the ionized electron. By contrast, the full TDSE solution displays significant deviations from the SFA predictions indicating a much more complex variation of the energy dependence interference phase $\delta(E)$. Even at relatively large energies above the threshold ($\sim 20$ eV), no clear indication for the convergence towards the SFA limit as assumed in previous analyses \cite{Zipp,Song18} emerges. These strong variations of $\delta(E)$ and deviations from SFA  appearing in the TDSE results are the signature of simultaneous interaction of the escaping electron with both the atomic force field and the strong laser fields, in particular intermediate off-shell bound-bound and continuum-continuum (cc) transitions between field-dressed atomic states \cite{Zhang10}. Such contributions are absent in the SFA and the CVA. 

\begin{figure}[tbp]
\includegraphics[width=10cm]{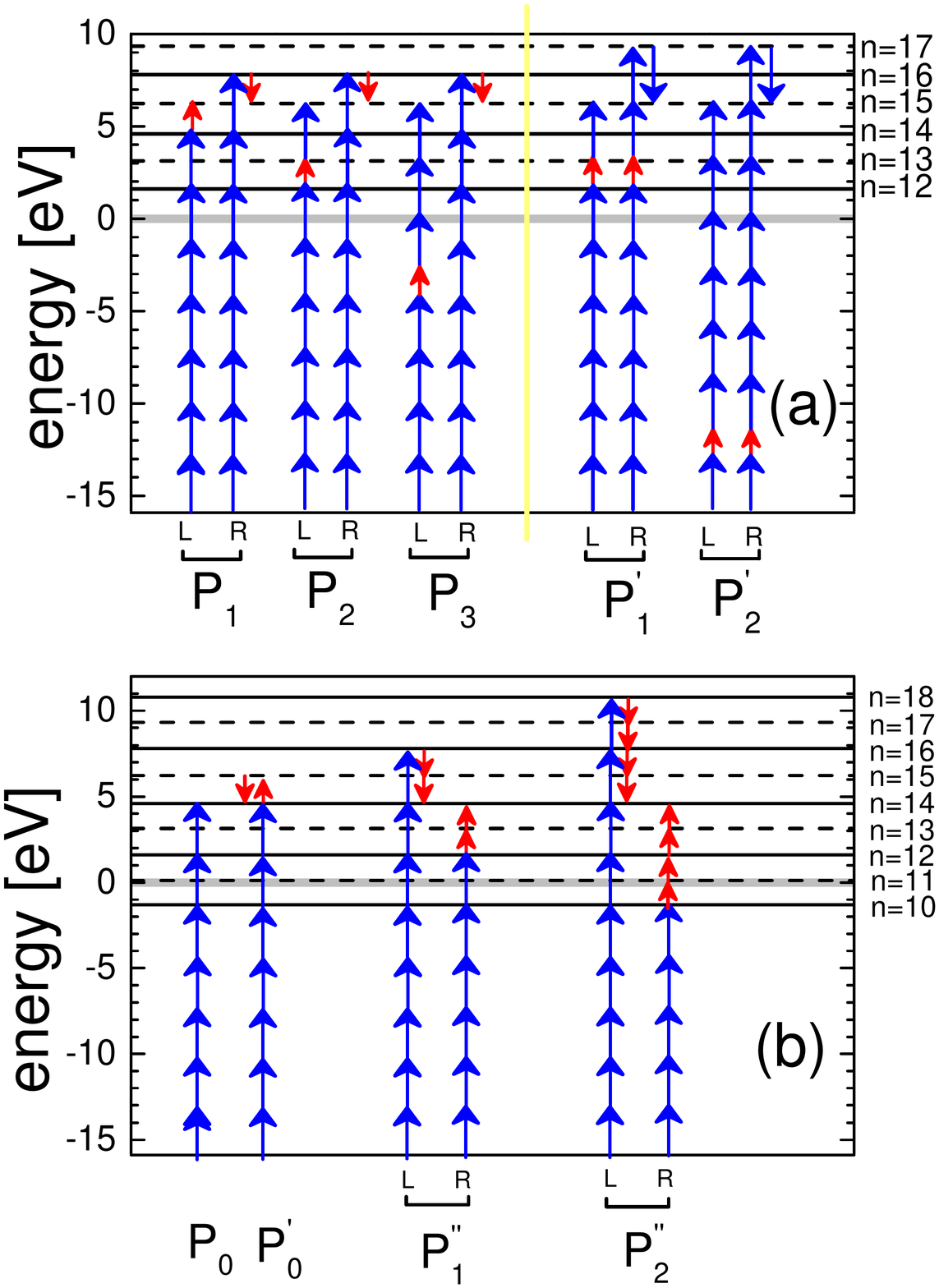}
\caption{Examples of pairs of quantum paths reaching the sideband energy
$E_{n}=E_{15}$ (a) or the ATI (or main) peak energy $E_{n}=E_{14}$ (b) for argon.
In (a), the set $P_i (i=1,...)$ features pairs, each with absorption (left L) or emission (right R) of one weak probe photon $\omega$ and the set $P_{i}^{\prime}$ features pairs with one additional absorption and emission of the strong pump photon (right R) compared to the direct path (left L) while both absorbing one weak probe photon $\omega$. The absorption of the probe photons may occur in the continuum ($P_1$, $P_2$ and $P_{1}^{\prime}$) or in virtual intermediate bound states ($P_3$ and $P_{2}^{\prime}$).
In (b), the direct ATI process can, to lowest order, interfere with path pairs $P_1$ involving absorption or emission of two $\omega$ photons, $P_2$ are examples of a process involving four $\omega$ photons. $P_{0}^{\prime}$ represents one contribution to the dressing of the ATI electron by the probe field.}
\label{scheme}
\end{figure}

Interpretation of the phase shift $\delta(E)$ of the forward (or backward) emission or asymmetry signal [Eq. (\ref{asym-cont})] requires a more detailed analysis of the interfering quantum paths. Key point is that in the present $\omega -2 \omega$ multi-photon strong-field interference (MPSFI) scenario a multitude of pathways contribute, a few of them shown for argon in Fig. \ref{scheme}, well beyond the subset invoked in the  analogy to the RABBITT protocol (Fig.~\ref{interference_sketch}a). This renders a quantitative analysis more challenging. 
For example, the side-band energy $E_n = E_{15}$ can be reached not only by the path pair $P_1$ (Fig.~\ref{scheme}a), which resembles the RABBIT protocol, but also by other path pairs with different sequences of absorption and emission events to the same first order in the weak probe field (e.g. $P_2$, $P_3$,...), or to different orders in the pump field (e.g. $P_1',P_2'$). The (virtual) intermediate states reached by the probe photon may involve continuum (e.g. $P_2,P_1'$) or bound states (e.g. $P_3,P_2'$). The latter are expected to be more important when the path proceeds via a bound-state resonance.

Multi-photon path interferences can be analyzed not only near the sidebands (Fig.~\ref{scheme}a) but also near ATI (or main) peaks (Fig.~\ref{scheme}b). For example,
at ATI energy $E_n=E_{14}$ the direct path $P_0$ from the initial state to the final state with $E_{14}$ via absorption of $7$ photons with frequency $2 \omega$ can interfere with a multitude of paths involving two probe photons $P_1''$ (Fig. \ref{scheme}b), which are of the same order in the weak field as the dressing of the ATI electron by the IR field $P_0'$. For a stronger probe field, even higher-order contributions may become important; examples of which involve the absorption or emission of 2 or 4 $\omega$ photons are shown  in Fig. \ref{scheme}b. It is important to realize that the set of paths in Fig. \ref{scheme} still do not fully reflect the complexity of the ensemble of contributing interfering paths as the angular momentum degree of freedom is omitted here for simplicity (see \cite{Bharti21}).
Each additional photon absorption or emission process leads to a branching of paths to multiply degenerate states of the same energy $E$ but different angular momenta $\ell \rightarrow (\ell+1,\ell-1)$. Consequently, for an initial state with angular momentum $\ell_i$ all partial waves $E$ within the interval $\left[\max(0,\ell_i-N),\ell_i+N\right]$ can be coherently populated at the final energy when the pulses are linearly polarized.

\section{Analytical model for quantum path interferences in multi-photon ionization}
\label{analytical_model}

\begin{figure}[tbp]
\includegraphics[width=10cm]{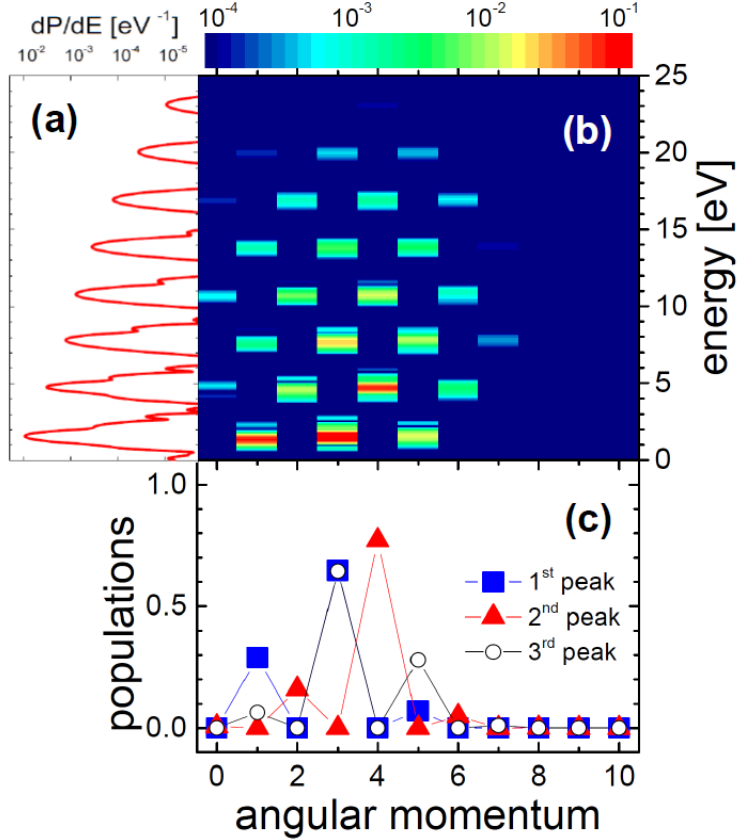}
\caption{ Energy spectrum and angular momentum distribution after strong-field ionization of argon by the one-color $2\omega$ field with the same parameters as in Fig. \ref{ar-spectrum}. (a) Photoelectron spectrum, (b) electron distribution as a function of the energy and angular momentum on a logarithmic scale covering three orders of magnitude, and (c) normalized  $p_{\ell}$ (integrated over energy) for the first three ATI peaks from threshold. 
}
\label{chessboard}
\end{figure}

In order to provide an intuitive guide towards interpreting the ionization phase shift $\delta(E)$ extracted from the quantum path interferences contributing to MPSFI we present a simplified analysis based on a (lowest-order) perturbative multi-photon description. Accordingly, the contribution of the $N$-photon absorption path (e.g., $P_0$ in Fig.~\ref{scheme}b) to electron emission in the $\theta$ direction following the absorption of $N$ photons of frequency $2\omega$ in the visible has the complex amplitude
\begin{equation}
	C\left(E_{2N},N\right)=\sum_{\ell} A_{N,\ell} \exp \left[ i\left(N \phi - N\frac{\pi}{2}-\ell \frac{\pi}{2} + \eta_{\ell} \left(E_{2N},F\right)  \right) \right] Y_{\ell}^0 \left(\theta\right).
\label{cplx_amplitude}
\end{equation}
In Eq.~\eqref{cplx_amplitude} $A_{N,\ell}$ is the modulus of the $N$-photon absorption amplitude and $\eta_{\ell}\left(E_{2N},F\right)$ is the atomic ionization phase at energy $E=E_{2N}$. In the weak field-limit this phase is expected to approach the one-photon atomic ionization phase at the same energy and angular momentum $\eta_{\ell} \left(E_{2N},F\rightarrow 0\right)=\eta_{\ell}\left(E_{2N}\right)$. However, in the present strong-field setting, deviations from this limit are expected.
The sum in Eq.~\eqref{cplx_amplitude} extends over all orbital quantum numbers fulfilling the inequality $\max\left[0,\ell_i -N \right] \leq \ell \leq \ell_{i}+N$. For estimating the phases in Eq.~\eqref{cplx_amplitude} we have used that each photon absorption or emission event contributes a phase $\pi/2$, each angular momentum change $\Delta \ell$ adds another $\Delta \ell \pi/2$, and each absorption of a $2\omega$ pump photon includes an additional relative phase $\phi$ of the pump field relative to the probe field [see Eq. (\ref{field})]. Applying now Eq.~\eqref{cplx_amplitude} to the left (L) path of pair $P_1$ (Fig.~\ref{scheme}a) contributing near the sideband energy $E_{2N+1}$, the combined amplitude for absorbing $N$ visible (V) $2 \omega$ photons followed by absorbing one NIR $\omega$ photon reads
\begin{eqnarray} \label{amplitude2}
C_{P_1,L}\left(E_{2N+1}\right)=&&\sum_{\ell,\sigma=\pm 1}  A_{N,\ell}^{\mathrm{V}} A_{1+,\sigma}^{\mathrm{NIR}}  \\  & &\exp\left\{i[N\phi-(N+1)\frac{\pi}{2}-(\ell+\sigma)\frac{\pi}{2}+\eta_{\ell}(E_{2N},F)+\varphi_{\ell+\sigma}^{cc,1+}(E_{2N},F)]\right\} Y_{\ell+\sigma}^0(\theta) \nonumber
\end{eqnarray}
with $\sigma=\Delta \ell = \pm 1$ the change in angular momentum due to the absorption of an additional NIR photon. 
$A_{1+,\sigma}^{\mathrm{NIR}}$ denotes the modulus and $\varphi_{\ell+\sigma}^{cc,1+}(E_{n-1},F)$ the corresponding additional phase of the absorption of one additional ($1+$) NIR photon.
It describes the continuum-continuum transition to the angular momentum sector $\ell+\sigma$ in the sideband reached by the absorption of $N$ photons of frequency $2\omega$ and one additional photon of frequency $\omega$, i.e., $n=2N+1$. In the perturbative limit, this phase is the analogue to the corresponding phase in RABBIT which depends, in general, on $\ell$ \cite{Fuchs20}. However, for probe fields beyond the perturbative limit, the continuum-continuum phase is expected to be dependent also on $F_{\omega}$. When both pump and probe fields are simultaneously present [(Eq. (\ref{field})], the phases will depend, in general, on the combined field $F$.
The corresponding expression for the right (R) of the path pair $P_1$ is accordingly given by 
\begin{eqnarray} \label{amplitude3}
	C_{P_1,R}\left(E_{2N+1}\right)=&& \sum_{\ell,\sigma=\pm 1} A_{N+1,\ell}^{\mathrm{V}} A_{1-,\sigma}^{\mathrm{NIR}} Y_{\ell+\sigma}^0(\theta) \\
	&& \exp\left\{i[(N+1)\phi-(N+2)\frac{\pi}{2}-(\ell+\sigma)\frac{\pi}{2}+\eta_{\ell}(E_{2(N+1)},F)+\varphi_{\ell+\sigma}^{cc,1-}(E_{2(N+1)},F)]\right\}  \nonumber
\end{eqnarray}
where $A_{1-,\sigma}^{\mathrm{NIR}}$ denotes the modulus and $\varphi_{\ell+\sigma}^{cc,1-}(E_{2(N+1)},F)$ the corresponding cc phase of the emission amplitude of an IR photon . Note that the range of $\ell$ included in Eq.~\eqref{amplitude3} is different from that in Eq.~\eqref{amplitude2} and includes $\max \left[0,\ell_i-(N+1)\right]\leq \ell\leq  \ell_i+N+1 $.
When, e.g., only the path pair $P_1$ in Fig. \ref{scheme}a is considered, the emission probability near the sideband $E=E_{2N+1}$ [Eq.~\eqref{hemispheres}] is now given by the coherent sum of Eq.~\eqref{amplitude2} and~\eqref{amplitude3},
\begin{equation}
S_{+(-)}(E_n,\phi)=\int_{0(-1)}^{1(0)}\mathrm{d}\cos\theta \:  \left|C_{P_1,L} \left(E_n\right) + C_{P_1,R} \left(E_n\right)\right|^2.
\label{hemispheres_model}
\end{equation}
The evaluation of Eq.~\eqref{hemispheres_model} can be drastically simplified by including only the dominant pathways along the so called ``yrast line'' well known from beam-foil spectroscopy \cite{BellNIM,BellPLA} or, equivalently, assuming that only the pathways preferred by the Fano propensity rule \cite{Fano85,Busto19} are realized.
Accordingly, each photoabsorption leads predominantly to an increase $\left(1+\leftrightarrow \sigma=1 \right)$ and photoemission to a decrease  $\left(1-\leftrightarrow \sigma=-1 \right)$ by one unit of angular momentum. Including only these dominant paths eliminates the summation over $\ell$ and $\sigma$ in Eqs.~\eqref{amplitude2} and~\eqref{amplitude3}.
We note that this approximate selection rule is only applicable to resonant bound-bound or continuum-continuum transitions but not to tunneling or above-threshold ionization. For ATI peaks close to threshold (th), the dominant $\ell$ values are delimited by \cite{Arbo2015b}
\begin{equation}
	\ell \leq \ell_{\mathrm{th}} \leq \left(2  Z_{T} \alpha \gamma\right)^{1/2} = \left(2 \sqrt{2} Z_{T} \sqrt{\frac{N_{\mathrm{th}}}{2 \omega}}\right)^{1/2}
\label{l_max}
\end{equation}
where $\alpha$ is the quiver amplitude, $\gamma$ the Keldysh parameter of the laser field with frequency $2\omega$, and $N_{\mathrm{th}}$ the minimum number photons of frequency $2\omega$ required to reach the continuum ($N_{\mathrm{th}}=6$ for argon). Accordingly, our TDSE calculations yield $f$ waves as dominant partial waves near threshold, which is very close to the upper bound predicted by Eq.~\eqref{l_max} $\ell_{\mathrm{th}}=4$ and well below the prediction for the yrast line (or propensity rule \cite{Fano85,Bertolino20}) $\ell_i+N_{\mathrm{th}}=7$ as depicted in Fig. \ref{chessboard}. The partial wave content of the first ATI peak above threshold and starting point of the further spread in angular momentum is thus centered at lower values of $\ell \leq \ell_{\mathrm{th}}$.
The evolution of the partial wave distribution $p_{\ell}$ to higher partial waves with increasing ATI peak is discernible (Fig. \ref{chessboard}c).
The first ATI peak exhibits a dominant angular momentum of $\ell_{\mathrm{th}}=3$, whereas for the second ATI peak the dominant angular momentum is $\ell=4$.
The combined contribution of the $d$ and $g$ waves of the second ATI peak produces a dominant $f$ wave ($\ell=3$) for the third ATI peak but with an appreciable $\ell=5$ contribution, i.e., $p_5 \simeq 0.5 p_3$.
Applying the approximate propensity rule to Eqs.~\eqref{amplitude2},~\eqref{amplitude3}, and ~\eqref{hemispheres_model} yields, e.g.,
\begin{eqnarray}
S_+(E_{2N+1},\phi) &=& \int_0^{1} \mathrm{d} \cos\theta \left\{ (A_{N,\ell}^{\mathrm{V}})^2 (A_{1+}^{\mathrm{NIR}})^2 (Y_{\ell+1}^0(\theta))^2 + (A_{N+1,\ell+1}^{\mathrm{V}})^2 (A_{1-}^{\mathrm{NIR}})^2 (Y_{\ell}^0(\theta))^2 \right.  \nonumber \\
& + & 2 A_{N,\ell}^{\mathrm{V}} A_{N+1,\ell+1}^{\mathrm{V}} A_{1+}^{\mathrm{NIR}} A_{1-}^{\mathrm{NIR}} Y_{\ell+1}^0(\theta) Y_{\ell}^0(\theta) \label{for-SB2} \\
& \times & \left. \cos\left[ \phi + \eta_{\ell+1}(E_{2(N+1)},F) - \eta_{\ell}(E_{2N},F) + \varphi^{cc,1-}_{\ell}(E_{2(N+1)},F) - \varphi^{cc,1+}_{\ell+1}(E_{2N},F) \right] \right\} . \nonumber
\end{eqnarray}
with an analogous expression for $S_-\left(E,\phi\right)$. 
Consequently, the asymmetry $A(E=E_{2N+1},\phi)$ given by Eq. (\ref{asymmetry}) is proportional to
\begin{eqnarray}
A(E_{2N+1},\phi) & \sim & S_+(E_{2N+1},\phi) - S_-(E_{2N+1},\phi) \nonumber \\ 
& \sim & 2 A_{N,\ell}^{\mathrm{V}} A_{N+1,\ell+1}^{\mathrm{V}} A_{1+}^{\mathrm{NIR}} A_{1-}^{\mathrm{NIR}} \int_0^{1} \mathrm{d} \cos\theta Y_{\ell+1}^0(\theta) Y_{\ell}^0(\theta) \label{asym-SB2} \\
& \times & \cos\left[ \phi + \eta_{\ell+1}(E_{2(N+1)},F) - \eta_{\ell}(E_{2N},F) + \varphi^{cc,1-}_{\ell}(E_{2(N+1)},F) - \varphi^{cc,1+}_{\ell+1}(E_{2N},F) \right] . \nonumber \label{asym-SB}
\end{eqnarray}
Comparison with Eq. (\ref{asym-cont}) yields now an explicit analytic but approximate expression of the phase delay between the two paths of the pair $P_1$ (Fig.~\ref{scheme}a)
\begin{equation}
\delta(E_{2N+1})\simeq \eta_{\ell}(E_{2N},F) - \eta_{\ell+1}(E_{2(N+1)},F) + \varphi_{\ell+1}^{cc,1+}(E_{2N},F) - \varphi_{\ell}^{cc,1-}(E_{2(N+1)},F) .
\label{phaseshift1}
\end{equation}
In the limit where all contributions to the phase of the wavepacket due to the interplay with the atomic force field and the laser field can be neglected, $\delta (E_{2N+1}) \approx 0$, Eq.~\eqref{asym-SB} reduces to 
\begin{equation}
A(E_{2N+1},\phi) \propto S_+(E_{2N+1},\phi) - S_-(E_{2N+1},\phi) = C \cos\phi
\label{SFA_SB}
\end{equation}
which agrees with the result in the SFA approximation first given by Zipp \textit{et al.} \cite{Zipp}.

A similar analysis for a pair of paths contributing to the asymmetry near the ATI energy $E_n$, where now $n=2N$, taking into account only interference between the direct ATI path $P_0$ and the path $P_1''(R)$ (Fig.~\ref{scheme}b) involving absorption of two NIR photons yields
\begin{eqnarray}
A(E_{2N},\phi) & \sim & S_+(E_{2N},\phi) - S_-(E_{2N},\phi) \nonumber \\
& \sim & A_{N,\ell}^{\mathrm{V}} A_{N-1,\ell-1}^{\mathrm{V}} A_{2+}^{\mathrm{NIR}}  \int_0^{1} \mathrm{d} \cos \theta Y_{\ell+1}^0(\theta) Y_{\ell}^0(\theta) \label{asym-ATI2} \\
& \times & \cos\left[ \phi + \pi + \eta_{\ell}(E_{2N},F)-\eta_{\ell-1}(E_{2(N-1)},F) - \varphi_{\ell+1}^{cc,2+}(E_{2(N-1)},F) \right] . \nonumber \label{asym-ATI}
\end{eqnarray}
with $A_{2+}^{\mathrm{NIR}}$ $\left(\varphi_{\ell+1}^{cc,2+}\right)$ the modulus (phase) of the two-photon transition amplitude from the ATI peak at $E_{n-2}$ with $\ell-1$ to $\left(E_n,\ell+1\right)$. 
Consequently, the phase delay between these two paths $\delta(E)$ is given by
\begin{equation}
\delta(E_{2N}) \simeq -\pi - \eta_{\ell}(E_{2N},F) + \eta_{\ell-1}(E_{2(N-1)},F) + \varphi_{\ell+1}^{cc,2+}(E_{2(N-1)},F) . 
\label{phaseshift2}
\end{equation}
In the limit that all atomic force field and laser field effects on the phase delay can be neglected, the SFA limit would emerge as 
\begin{equation}
S_+(E_{2N},\phi) - S_-(E_{2N},\phi)= A \cos\left(\phi + \pi \right) ,
\label{SFA_ATI}
\end{equation}
which results, indeed, in a phase jump of $\pi$ between the sidebands [Eq.~\eqref{SFA_SB}] and the ATI peaks [Eq.~\eqref{SFA_ATI}] in agreement with our numerical results (Fig. \ref{models-delays-cont}). Consequently, the deviations observed in the TDSE simulation and CVA simulations from these SFA limit are an unambiguous signature of the interplay between the atomic force field and laser fields in the atomic ionization phases. It should be emphasized that the TDSE results include all paths contributing to the multi-photon strong field interference for photoelectron well beyond the simple ``two-path double-slit'' model [Eq.~\eqref{asym-SB} and \eqref{SFA_ATI}] explicitly treated above. 

\begin{figure}[tbp]
\includegraphics[width=10cm]{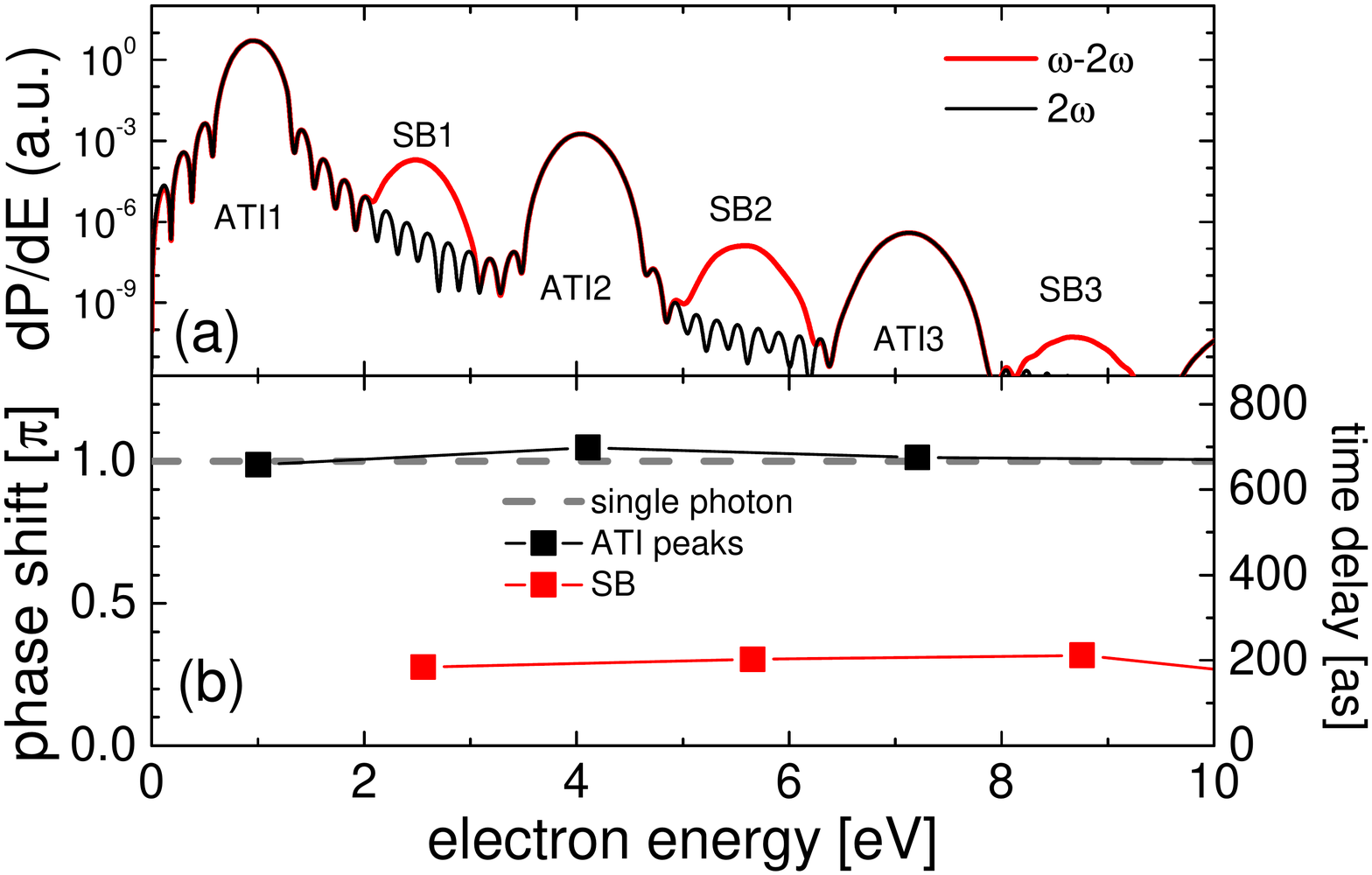}
\caption{(a) Electron spectra for a Yukawa potential [Eq.~\eqref{yukawa-potential}] with $a=4$ and $b=0.629$ calculated for one-color 2$\omega$ (black line) and two-color $\omega - 2\omega$ (red line) laser fields with $\phi=0$. (b) Phase delays $\delta(E)$ in units of $\pi$ calculated from the asymmetry $A(E,\phi)$ integrated over hemispheres [see Eq.~\eqref{hemispheres}]. For reference we also convert the phase delay into a time delay [Eq.~\eqref{delay-cont}] (right side axis). The laser intensities are $I_{2\omega} = 10^{11}$ W/cm$^2$ and $I_{\omega} = 5 \times 10^{8}$ W/cm$^2$. Other laser parameters are the same as in Fig. \ref{ar-spectrum}.}
\label{shallow-yuka}
\end{figure}

The two-path model can provide guidance as to which information can be extracted from MPSFI spectra. For example, the phase contributions $\eta$ and $\phi^{cc}$ will be, in general, dependent on the field strengths $F_{2\omega}$ and $F_{\omega}$ in a strong-field $\omega-2\omega$ scenario fundamentally different from the standard RABBIT protocol. Moreover, while the resulting phase delay $\delta (E)$ is a continuous function of $E$ (see Fig.~\ref{models-delays-cont}) the mapping of a phase delay onto a time delay according to Eq.~\eqref{delay-cont} depends on the specific position within the spectrum. Near sideband energies $E_{2N+1}$, Eq.~\eqref{phaseshift1} has the appearance of a finite difference approximation as implied by Eq. (\ref{delay-cont}) and can thus be used to extract approximate time delays $\tau=\delta (E_{2N+1}) / 2 \omega$. 
Near ATI peaks [Eq. (\ref{phaseshift2})], such interpretation in terms of a finite-difference approximation fails as the difference involves now different interfering zero- and two-IR photon paths.
Moreover, when all path pairs are included, a sum over many path pairs each of which giving rise to terms of the form [Eq.~\eqref{asym-SB}] for sidebands and  of the form [Eq.~\eqref{asym-ATI}] for ATI peaks will contribute to $A(E,\phi)$ rendering the extraction of a spectral derivative for a specific phase difficult.
Only in cases where one path pair strongly dominates, in particular the pair $P_1$ for the sideband, approximate EWS time delays for a given partial wave can be unambiguously assigned. With this caveat in place, we also give $\tau(E)$ in Figs.~\ref{shallow-yuka},~\ref{ar-delays}, and ~\ref{delays-vs-I2} for illustrative purposes.

Before comparing simulations with experimental data, we illustrate the partial-wave path-interference structure for a strongly simplified model system in which the number of contributing paths and, thus, the complexity of the ionizing process is drastically reduced. We consider an electron bound by a Yukawa potential [Eq. (\ref{yukawa-potential})] with parameters ($a=4$, $b=0.629$) chosen such that a single $2\omega$ photon is sufficient to reach the continuum and the shallow potential supports only one 1s-like bound state with $E_{1s}=-0.08$. Consequently, the energetic position of the first ATI peak coincides in this case with the position of the standard photoionization peak. For later reference we note that the screening length of this potential ($a=4$) is sufficiently large as to include, despite being asymptotically short-ranged, some Coulomb-laser coupling (CLC) or cc phase contributions \cite{Nagele2013}. Moreover, we choose the intensities of the fields sufficiently low ($I_{2\omega}=10^{11}$ W/cm$^2$, $I_{\omega}=5\times10^{8}$ W/cm$^2$) to be strictly in the perturbative multi-photon regime.
The photoelectron spectrum in both the presence and absence of the weak probe field are displayed in Fig. \ref{shallow-yuka}a. Turning on the $\omega$ field creates the side bands, as expected, while the ATI peaks remain largely unaffected by the probe field.
The absorption of a single V ($2\omega$) photon from the bound $1s$ initial state ionizes the model atom creating a $p-$wave electron of energy corresponding to the first peak (ATI1).
The second peak (ATI2) results from the absorption of two-V($2\omega$) photons, and is composed of the superposition of \textsl{s} and \textsl{d} waves due to the selection rule of angular momentum $\Delta \ell = \pm 1$. We have determined the angular momentum composition of ATI2 to contain $9.8\%$ of \textsl{s} character and $90.1\%$ of \textsl{d} character consistent with the propensity rule invoked above.
The lowest sideband SB1 between the first $2\omega$ photoionization peak ATI1 and the second peak ATI2 can be reached by either absorption of two photons [one V:($2\omega$) and one NIR:($\omega$)] or absorption of two V ($2\omega$) photons and emission of one NIR ($\omega$) photon.
For the first sideband SB1 the angular momentum composition is given by $9.4\%$, $0.8\%$, and $89.8\%$ for the $s$, $p$, and $d$ states, respectively. The population of $s-$ and $d-$partial waves in SB1 is close to that of ATI2 also in line with the propensity for two-photon absorption irrespective of the different frequencies involved.
This distribution indicates the dominance of the one-V($2\omega$)-one-NIR($1\omega$) absorption path to the SB1 over the two-V($2\omega$) absorption and one-NIR($1\omega$) emission path in the perturbative regime, which is expected since the latter path involves one more photon from a weak field than the former and, consequently, is a higher-order photoionization process. However, the latter path provides a small but crucial contribution giving rise to a non-vanishing $\phi$ dependent contribution from which the phase delay $\delta(E)$ can be extracted (Fig. \ref{shallow-yuka}b).
 
Remarkably, whereas $\delta(E)$ near the ATI peaks closely follows the SFA predictions $\delta(E_{n}) \simeq \pi$ [Eq.~\eqref{SFA_ATI}], near the sideband peaks strong deviations can be observed in Fig. \ref{shallow-yuka}b. For the first sidebands for which this phase could be reliably extracted we find $\delta(E_{n}) \simeq 0.3\pi$.
For reference we also convert the phase delay $\delta (E)$ into an EWS-type time delay following Eq.~\eqref{delay-cont} and find for the sideband, within a fairly small energy window ($3$ eV $\leq E \leq$ $10$ eV), an almost energy-independent time delay of about $\tau \approx 200$ attoseconds. Using the approximate expressions [Eqs.~\eqref{SFA_SB} and~\eqref{SFA_ATI}] for a qualitative analysis of the two-path interference these results suggest that the phase delay near the ATI peaks is strongly dominated by the SFA contribution ($\sim \pi$) corresponding to a time delay of $660$~as while atomic field corrections play only a minor role.
By contrast, near the sideband peaks the phase differences induced by the atomic-field  $\eta_{\ell+1}(E_{2(N+1)})-\eta_{\ell}(E_{2N})+\varphi_{\ell}^{cc,1-}-\varphi_{\ell+1}^{cc,1+}$ are clearly visible. We note that the presence of a non-vanishing contribution to the phase delay by the one-photon continuum-continuum transition $\varphi^{cc,1\pm}$ for the Yukawa potential is consistent with the fact that with increasing screening length ($a=4$ in the present case) an increasing part of the full long-range Coulomb-laser coupling is restored \cite{Nagele2013}. Therefore, we can use Eq. (\ref{phaseshift1}) to estimate this contribution to the sideband phase delay as
\begin{equation}
\varphi_{\ell+1}^{cc,1+}(E_{2N}) - \varphi_{\ell}^{cc,1-}(E_{2(N+1)})
\simeq \delta(E_{2N+1}) + \eta_{\ell+1}(E_{2(N+1)}) - \eta_{\ell}(E_{2N}),
\label{cc}
\end{equation}
where we have dropped the label $F$ because we consider the perturbative limit ($F \rightarrow 0$). The atomic ionization phases $\eta_{\ell}$ can be obtained by the one-photon atomic ionization phase in a partial-wave expansion for the Yukawa potential. By using Eq. (\ref{cc}), we estimate the cc phase contribution to SB1 as $\varphi_{2}^{cc,1+} - \varphi_{1}^{cc,1-} \simeq 0.45$, for SB2 as $\varphi_{3}^{cc,1+} - \varphi_{2}^{cc,1-} \simeq 0.8$, and for SB3 as $\varphi_{4}^{cc,1+} - \varphi_{3}^{cc,1-} \simeq 0.92$, corresponding to time delay contributions of approximately $11$, $19$, and $22$ as, respectively. These phase contributions could shed some light on how the Yukawa potential affects the cc contributions to the time delays. Besides, new studies on the holographic angular streaking of electrons by corotating ($\omega-2\omega$) fields suggest that nonadiabatic effects in the ionization process could be responsible for such difference of the time delay with respect to the strong-field approximation \cite{Eckart20,Trabert21}. The identification of non-adiabatic effects on time delays (included in the TDSE calculations) are beyond the scope of this paper. It is worth to mention that as the De Broglie's wavelength of the electron is longer than the screening length of the Yukawa short-range potential, classical or semiclassical simulations are not valid for the energy region shown in Fig. \ref{shallow-yuka}b.

\section{Comparison with experiment}
\label{results}

\begin{figure}[tbp]
\includegraphics[width=10cm]{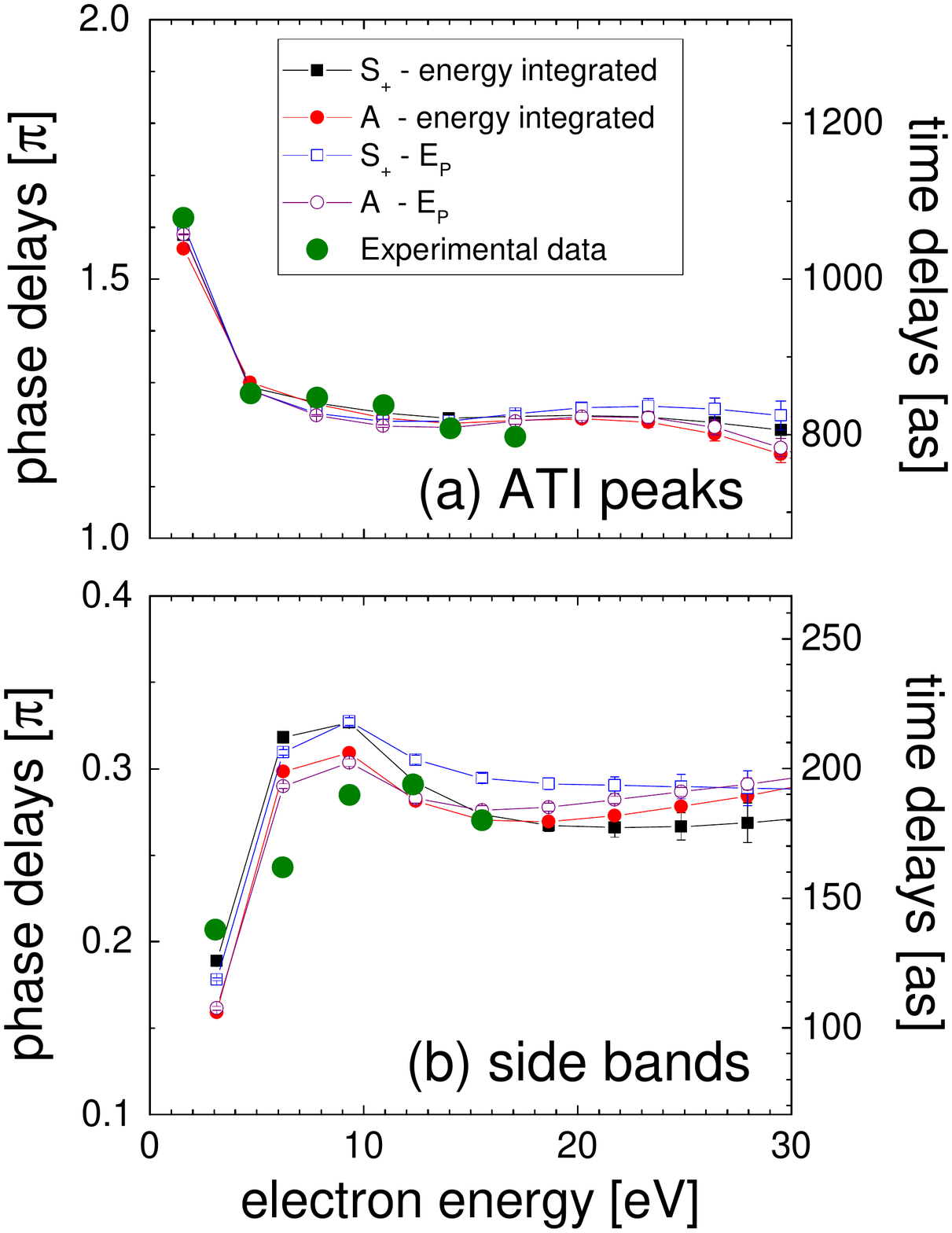}
\caption{ TDSE phase delays $\delta(E)$ calculated as a function of the emission energy for (a) ATI peaks and (b) sidebands for the same pulse parameters as in Fig. \ref{ar-spectrum}. Phase shifts extracted from data for forwards half spheres $S_+(E,\phi)$ (squares), and asymmetry $A(E,\phi)$ (circles) with integration over the energy window around each peak energy (full symbols) and at the energy peak only (open symbols). Full green dots correspond to experimental data by Zipp \textit{et al.} \cite{Zipp} normalized to the TDSE result at the highest sideband energy 
($\sim 15$eV).
}
\label{ar-delays}
\end{figure}

For a comparison with the experiment of Zipp \textit{et al.} \cite{Zipp} we extract the multi-photon ionization interference phase shifts $\delta(E)$ from the TDSE simulation (Fig. \ref{ar-spectrum}). In view of the rapid variation with the energy $E$ (Figs. \ref{ar-spectrum} and \ref{models-delays-cont}), we evaluate $\delta(E)$ not only at the ATI or sideband peaks $E=E_n$ [Eq. (\ref{Econs})] but integrate the spectrum over an energy window of width $\Delta E = 0.3 \omega$ centered around the peak. 
We show in Fig. \ref{ar-delays} fits to $\delta(E)$ for emission into forward hemisphere $S_+(E, \phi)$ [Eq. (\ref{hemispheres})] and for the asymmetry $A(E,\phi)$ [Eq. (\ref{asymmetry})].
While minor differences of the order of less than $0.05\pi$ between the different read-outs of $\delta(E)$ (via $S_+$ or $A$) appear, the overall trends observed are independent of the particular read-out protocol demonstrating that unambiguous information on the phase delay can be extracted.

For further analysis and interpretation of the results of Fig.~\ref{ar-delays}, two key points should be taken into account. First, the experimental data for $\delta(E)$ presented in \cite{Zipp} were relative and set to coincide with the SFA value ($\delta=0$) at the highest energy measured ($E=15$eV) (a similar renormalization was used in \cite{Song18}). However, we observe significant deviations in $\delta(E)$ from the SFA limit. Therefore, we instead renormalize the experimental data to the full TDSE result at the highest experimental energy in order to preserve this additional information on the absolute value of $\delta(E)$. Accordingly, in Figs. \ref{ar-delays}a and~b the experimental results are set to coincide with the TDSE phase shifts calculated by integration over the energy windows around the peaks and all angles in the forward hemisphere. Overall, the trend in the experimental data is well reproduced by the simulations. The sharp rise of the phase shift $\delta(E)$ for the first ATI peak seen close to threshold in both the experiment and simulations was recently interpreted in terms of transient trapping of the electron in Rydberg states by the $\omega - 2\omega$ field \cite{Song18}.

The second key feature is that the data in Fig.~\ref{ar-delays} were extracted at a moderately strong NIR probe field with $I_{\omega}=4 \times 10^{11}$~W/cm$^2$. For the standard RABBIT protocol or attosecond streaking field strengths $F_{\omega}$ of that order of magnitude were found to be weak enough to unambiguously extract atomic continuum-continuum or Coulomb-laser coupling delays which are independent of the particular value of $I_{\omega}$ in line with lowest-order perturbation theory \cite{PazourekRMP15}. However, in the present MPSFI scenario the influence of the probe field $F_{\omega}$ beyond a lowest-order perturbation theory must be considered.
\begin{figure}[tbp]
\includegraphics[width=10cm]{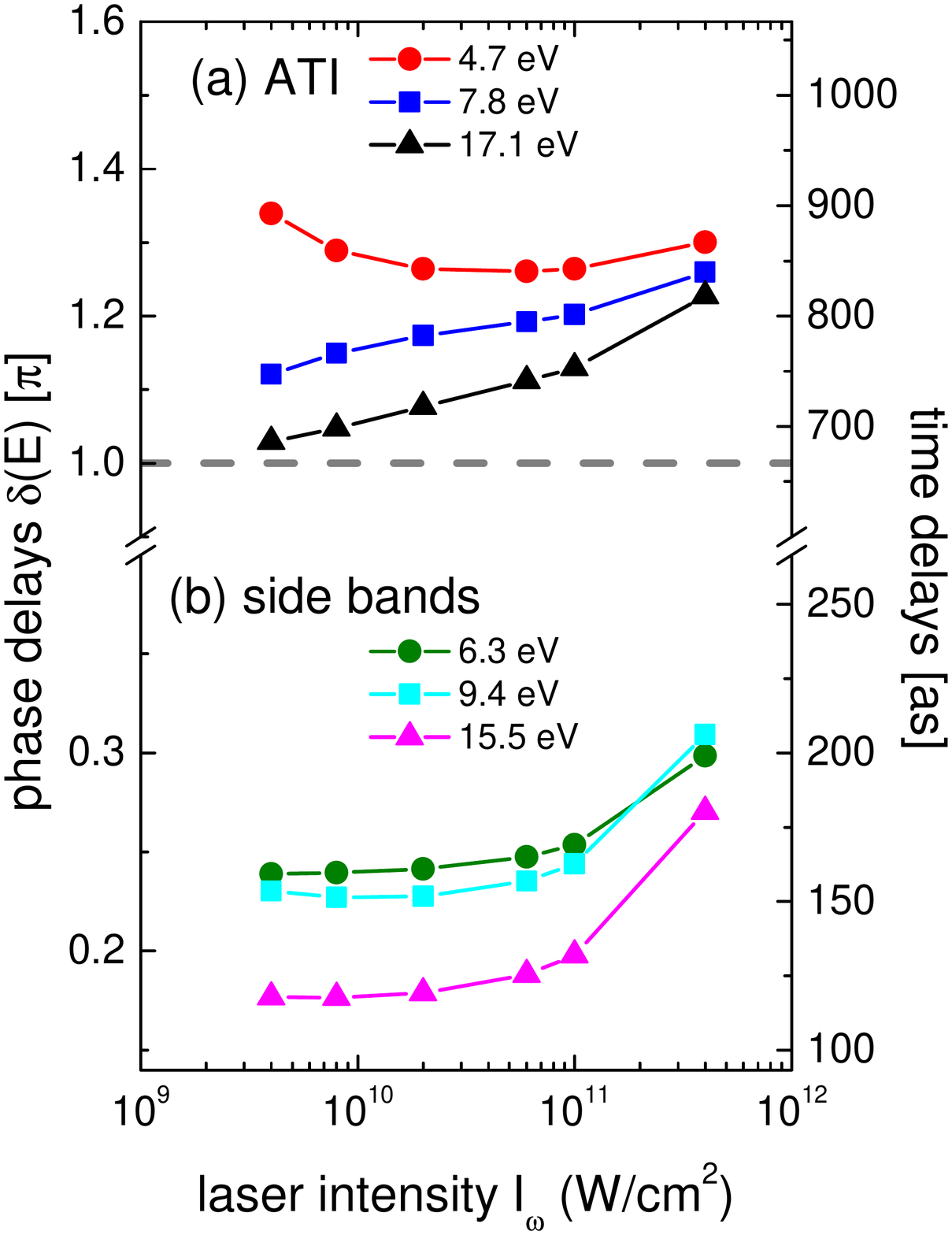}
\caption{Interference phase delay $\delta(E)$ as a function of the probe laser intensity $I_{\omega}$ extracted from asymmetry parameter integrated over hemispheres for three ATI peaks and three sidebands with energies as indicated. 
All other laser parameters are the same as in Fig. \ref{ar-spectrum}.
The horizontal dashed line corresponds to the strong-field limit for ATI phase shifts [$\delta(E)=\pi$], the SFA limit for the sidebands is $\delta(E)=0$ (not shown).
}
\label{delays-vs-I2}
\end{figure}
Indeed, exploring the variation of the extracted $\delta(E)$ at fixed pump intensity $I_{2\omega}$ as a function of the probe intensity $I_{\omega}$ (Fig. \ref{delays-vs-I2}) reveals a surprisingly strong dependence. The experimental value $I_{\omega}=4 \times 10^{11}$~W/cm$^2$ is obviously well beyond the lowest-order perturbative regime which precludes the  direct applicability of a RABBIT-type analysis.
For sideband peaks, phase shifts $\delta(E)$ appears to converge to the perturbative field-independent limit only for considerably lower fields $I_{\omega} \lesssim 10^{10}$ W/cm$^2$. These converged values differ, however, significantly from the SFA limit even at the highest energy measured ($E=15.5$ eV).
Near ATI peaks, variations are present even at such low intensities and the approach to converged field-independent values is not yet obvious. It appears that for the highest energies measured, e.g. $E=17.1$ eV and at the lowest probe field  $I_{\omega} \lesssim 10^{10}$ W/cm$^2$ the phase near the ATI peak may approach the SFA limit $\delta(E) \simeq \pi$.
It should be noted, however, that the interference contributions to ATI peaks, which are responsible for the phase shift $\delta(E)$, result from (at least) a two-photon absorption or emission event in the probe field ($P_1''$ as depicted in Fig. \ref{scheme}b), which becomes very weak at low $I_{\omega}$ rendering the phase extraction uncertain.
The non-negligible probe field dependence of the extracted MPSFI phase delays $\delta(E)$, also indicated in Eqs.~\eqref{phaseshift1} and~\eqref{phaseshift2} emerges as an important new feature, absent in standard RABBIT or streaking measurements, that remains to be explored, experimentally as well as theoretically.

\section{Concluding remarks}
\label{conclusions}
We have presented simulations and the first detailed analysis of the phase delays $\delta(E)$ in 
multi-photon ionization. They provide information on the differences in ionization phases among different pathways open in a $\omega - 2\omega$ scenario for atomic ionization. We show that $\delta(E)$ is determined by quantum path interferences between different sequences of photon absorption and emission events. In the SFA limit these phases are given by $\delta(E)=0$ at sideband energies and by $\delta(E)=\pi$ at the ATI peaks.
We find that the solutions of the time-dependent Schr\"{o}dinger equation predict phases strongly differing from these SFA limits even at relatively high electron emission energies. We relate these phase shifts to the interplay between the strong $\omega - 2\omega$ field and the atomic force field not accounted for by the SFA. We  also point out the intrinsic difficulties to relate the phase delays $\delta(E)$ to time delays in analogy to the standard RABBIT protocol for one-photon ionization.
A multitude of different interfering pathways provides obstacles for a straightforward extraction of a spectral derivative of the phase delay.
We have found strong variation of $\delta(E)$ with the intensities of the pump and probe fields.  Our analysis shows that further experimental insight into the multi-photon ionization phase delay $\delta(E)$ can be gained by exploring its variation with both $I_{2\omega}$ and $I_{\omega}$.

\begin{acknowledgments}

This work was supported by CONICET PIP0386, PICT-2016-0296 PICT-2017-2945 and PICT-2016-3029 of ANPCyT (Argentina), Austria-Argentina
collaboration AU/12/02, by the FWF special research
programs SFB-041 (ViCoM), and doctoral programme
DK-W1243 (Solid4Fun), and by the European COST Action CA1822.
The computational results presented have been achieved using the Vienna Scientific Cluster (VSC).
\end{acknowledgments}

\end{document}